\documentclass[twocolumn,final]{svjour3}          % twocolumn
\smartqed  % flush right qed marks, e.g. at end of proof
\usepackage{amstext}
\usepackage{graphicx}
\usepackage{fix-cm}
\usepackage{mathptmx}      % use Times fonts if available on your TeX system
\journalname{Microgravity Science and Technology}
\begin{document}

%--------------------------------------------------------------------------------------------------

\title{Drop tower setup for dynamic light scattering in dense gas-fluidized granular media}


\author{Philip Born \and Johannes Schmitz \and Michael Bu\ss mann \and Matthias Sperl}

\institute{P. Born, J. Schmitz, M. Bu\ss mann, M. Sperl \at
              Institut f\"ur Materialphysik im Weltraum, Deutsches Zentrum f\"ur Luft- und Raumfahrt, 51170 Cologne, Germany  \\
              Tel.: +49-2203-601-4048\\
              Fax: +49-2203-61768\\
              \email{philip.born@dlr.de}                 %  \\
}

\date{Received: date / Accepted: date}

\maketitle
%\linenumbers

%--------------------------------------------------------------------------------------------------

\begin{abstract}
Investigation of dynamics in dense granular media is challenging. Here we present a setup that facilitates gas fluidization of dense granular media in microgravity. The dynamics is characterized using diffusing wave spectroscopy. We demonstrate that agitated granular media reach a steady state within fractions of a second in drop tower flights. The intensity autocorrelation functions obtained in microgravity show a remarkable dependence on sample volume fraction and driving strength. A plateau in correlation emerges at low volume fractions and strong driving, while correlation decays only very slowly but continuously at high packing fractions. The setup allows to independently set sample volume fraction and driving strength, and thus extends the possibilities for investigations on dynamics in dense granular on ground.
\end{abstract}

%--------------------------------------------------------------------------------------------------

\section{Introduction}
\label{intro}

The processes in granular media are strongly affected by gravity. The rapid settling of the particles and dissipation of all kinetic energy is an every day experience. The interplay among dissipation, settling and anisotropic agitation leads to density inhomogeneities and convection even if the particles are agitated into a steady state \cite{Poschel2000,Lohse2007}. Microgravity thus opens a new parameter space for studies on granular media. The absence of symmetry-breaking by gravity facilitates the use of isotropic agitation mechanisms in dilute granular media \cite{Meyer2013,Stannarius2013} and evaluation of processes in dust clouds of the nascent solar system \cite{Blum2000}. 

Investigation of dense granular media is particularly challenging. The collisional dynamics on a local scale may be very fast and requires high temporal and spatial resolution. The granular medium itself is highly opaque, unless index-matched by the surrounding liquid, what limits the optical investigation methods used for dilute media to investigations of particles at the surface \cite{Losert2013}. Attempts to investigate dense granular media so far tested index-matching for investigations on phenomena in static granular packings and X-ray radiography in combination with tracer particles for investigations in dynamic dense media \cite{Yu2014}. These methods were successfully applied and promise to deliver insights into specific scientific questions, but suffer from inherent limitations: dynamics is either highly damped by the index-matching fluid, or the measurements suffer from the limiting trade-off between spatial or temporal resolution of x-ray detectors.

Here, we use dynamic light scattering for measurements on dense agitated granular media. The evaluation methodology for dynamic light scattering had been extended to the regime of strong multiple scattering within the sample (DWS - diffusing wave spectroscopy) \cite{Pine1990}. DWS relies on measuring the speckle intensity fluctuations of coherent light diffusively transmitted through or reflected from the sample. The intensity fluctuations are further evaluated by the intensity autocorrelation function. From the intensity autocorrelation function information on particle displacement, granular temperatures and possible localization of the particles could be drawn when Gaussian intensity statistics are present and the light propagation mechanisms are known. Dynamics on nanoseconds and nanometer scales can be investigated with modern hardware correlators and using visible laser light \cite{Goldman2006,Xie2006}.

The setup presented in this work is designed for characterization of the dynamics in granular media close to the jamming transition. This aims at investigating the puzzling relation among jamming in granular media and the glass transition in thermal systems \cite{Sperl2012}. The generation of a steady state in such a dense granular medium is also a non-trivial task. Previous works on dense granular media relied on the g-jitter on parabolic flights \cite{Yu2014} or vibrating sample cells \cite{Hou2012}. These methods suffer from the dissipative collisions among the particles. Only a small fraction of the particles is directly agitated by the sample container walls, from where the kinetic energy has to be transferred by collisions. As energy is lost in each collision, gradients in dynamics or even granular clustering far from the vibrating surfaces occur \cite{Evesque2010}. Also, only few agitation directions can be realized using vibrating sample cells. We consequently adapt the method of gas-fluidization for our microgravity setup \cite{Grace2006}. Gas-fluidization requires directing a gas stream through the granular packing. This provides the advantages that in principle every particle can individually take up kinetic energy by the gas drag and a high number of agitation directions for homogeneous dynamics can be realized at least in microgravity. This comes at the cost of not providing undisturbed collisional dynamics of the particles within the gas stream in the granular medium.

We discuss in the following some aspects of light scattering and gas fluidization relevant for the design of the setup. Then the design and the realization of the setup for drop tower flights is presented. Finally, first measurements using this setup in microgravity are discussed. The results show the feasibility of studies with granular media close to jamming using gas-fluidization and dynamic light scattering.

%--------------------------------------------------------------------------------------------------

\section{Light scattering in microgravity}

\begin{figure}
	\centering
	\includegraphics[width=0.45\textwidth]{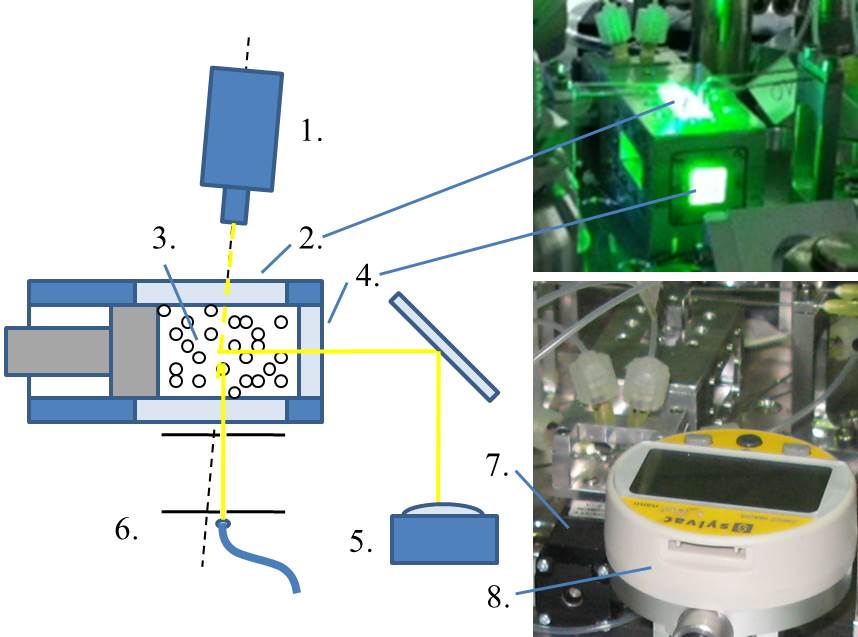}
	\caption{Sketch of the light scattering configuration: A laser (1.) shines through the top window (2.) of the sample cell into the granular medium (3.). The scattered light is either detected through the side window (4.) by the overview camera (5.) or after transmission through the medium and the bottom window by a collimator and a fiber (6.). The axis of the laser is at an angle of 5$^{\circ}$ relative to the axis defined by the optical fiber and the collimating pinholes. The upper right image shows a side view of the illuminated sample cell. The lower right image shows a rear side view of the sample cell with stepper motor (7.) and distance gauge (8.).}
	\label{fig:2} 
\end{figure}

DWS relies on measuring the temporal fluctuations in intensity $I(t)$ of coherent light which propagated diffusively through a sample. From the time-series analysis of the intensity fluctuations the mean-squared displacement of the scattering centers within the sample can be obtained \cite{Pine1990}. A diffusion-like propagation of the light can only be obtained if the path-lengths of the light within the sample are long compared to the randomization length of the light direction, or transport mean free path,  $l^{\ast}$ of the sample. The intensity of the light decays exponentially with the propagation distance, with the decay constant absorption length $l_{a}$. Both characteristic lengths $l^{\ast}$ and $l_{a}$ were found to be a few millimeters for submillimeter sized dielectric particles \cite{Ricka1996}. The measured light intensities are rather insensitive to the scattering angle in the regime of diffusion-like light transport, in contrast to conventional dynamic light scattering with single scattering propagation within a sample. Thus only the two fundamental scattering geometries of light being backscattered or being transmitted through the sample are usually considered for DWS \cite{Pine1990}.

We design the setup to facilitate DWS measurements in drop tower flights initiated by catapult starts and for sounding rocket flights. This imposes the constraints that the overall dimensions of the setup are limited, that the setup has to withstand the strong accelerations during start and landing, and that the measured intensity of diffusively propagated light is high to ensure a good signal-to-noise ratio within the limited measurement time. 

We chose a transmission geometry for our measurements, where light has to propagate through the whole sample cell, thus probes a large fraction of the sample volume and is likely to fulfill the requirement for diffusion-like propagation. Measurements in backscattering can be expected to be affected by a large fraction of light which is statically reflected from sample cell windows without an index-matching bath. The integration of such an index-matching bath to suppress specular reflections from the sample cell is constrained by the requirements of overall small dimensions and mechanical stability. An index-matching bath also complicates integration of the required agitation mechanism for fluidized granular media. Additionally, measurement in backscattering mainly probe a volume close to the sample cell window, where light scattered by the particles has a high chance of being scattered back out of the sample cell.  

Fig.~\ref{fig:2} shows the realized sample cell and light scattering scheme. We chose the sample cell size to be only few times the randomization length to ensure a sufficient transmitted intensity. We also omit polarizing optics and thus loose speckle contrast for the sake of high transmitted intensity. We introduce a short collimator with length of 3~cm and pinhole sizes of 2~mm under the bottom window of the sample cell and tilt the laser (CNI MXL-III-532) by 5$^{\circ}$ in order to protect the detectors. Possible accumulation of the particles at some side of the sample cell during take-off or landing cannot be excluded when measuring at low volume fractions, which would expose the detectors directly to the laser without further countermeasures.

The light is fed into a single-mode fiber (Thorlabs) and guided into a beamsplitter (Sch\"after+Kirchhoff FBS-532-Y) and two avalanche photodiodes (ID Quantique id100-MMF50-ULN). The signal is evaluated using a hardware correlator (ALV 7002/USB-25) by cross-correlation of the two detector signals to suppress afterpulsing effects. The hardware correlator provides a fast count rate trace with a time resolution of 200~$\mu$s and a correlation function with 25~ns sampling time.

The light intensity signal reaching the correlator is a function of the laser intensity, the density and configuration of the granular medium in the sample cell, and the detection efficiency of the detection setup, consisting of mono-mode fiber, beamsplitter and two avalanche photodiodes. All three factors could independently lead to intensity fluctuations, especially during a microgravity experiments initiated by a catapult or a rocket with strong accelerations. We use the first flights to check how the catapult start affects laser intensity and detection efficiency and thus influences the measured intensity fluctuations.

%--------------------------------------------------------------------------------------------------

\section{Gas-fluidization in microgravity}
\label{fluidization}

\begin{figure}
	\centering
	\includegraphics[width=0.45\textwidth]{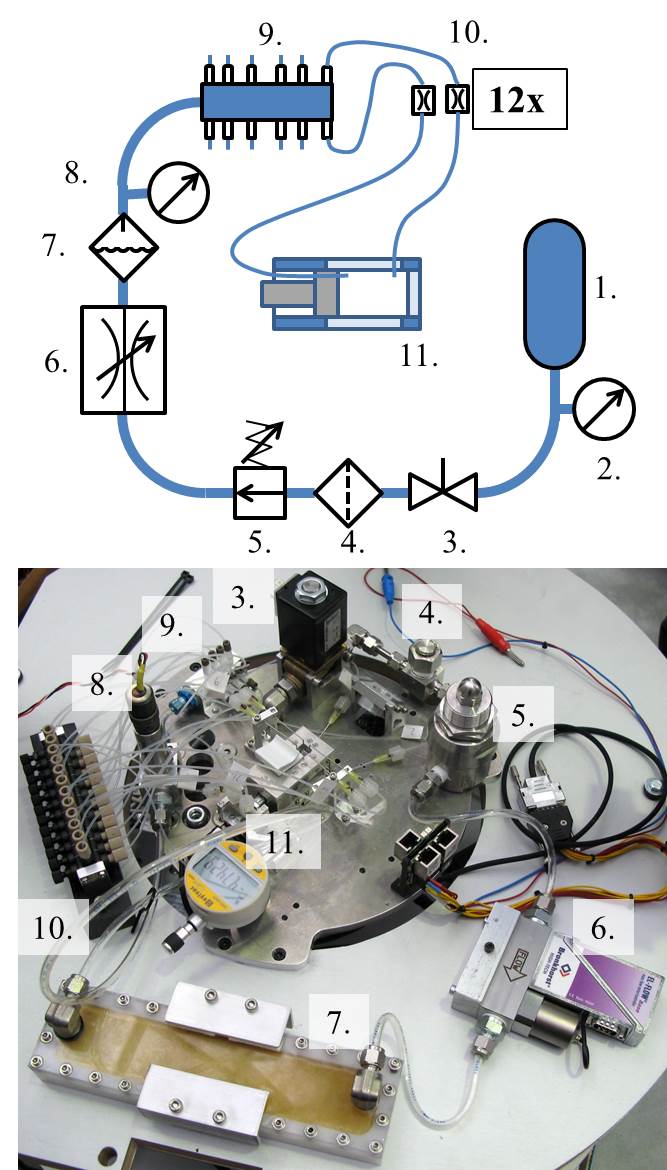}
	\caption{Scheme of the middle base plate with the gas agitation system and the realization for the drop tower experiments: 1. gas reservoir (mounted under the base plate), 2. pressure sensor, 3. shut-off valve, 4. filter, 5. pressure reducing valve, 6. mass flow control, 7. humidity chamber, 8. pressure sensor, 9. gas distributor, 10. twelve needle valves, 11. sample cell with twelve gas inlets.}
	\label{fig:3} 
\end{figure}

On ground, gas-fluidization requires that a gas stream is forced upwards through a packing of particles. The packing expands and the particles start to move when the pressure drop across the packing balances the weight of the packing and the interparticle cohesion. The granular medium then behaves in some aspects like a fluid, as the angle of repose becomes zero and the upper surface stays horizontal \cite{Castellanos2005}. The minimal gas velocity $U_{min}$ that has to be reached to gain a fluidized state is determined by gravity and particle cohesion \cite{Castellanos1999}. 

The need for directional forcing the gas through granular packing against the gravity direction ceases in microgravity experiments. We guide gas through 12 inlets with different directions into the granular medium. 8 inlets ensure that no dead corners are formed close to the windows of the sample cell  for the light scattering measurements. 4 additional inlets agitate the particles through the piston to prevent calm volumes when the piston is retracted (see Fig.~\ref{fig:3}). 

A controlled agitation strength is achieved by adjusting the mass flow of the gas stream. We reduce the pressure from 8~bar in the gas reservoir to 1.5 bar at the mass flow controller using a pressure reducing valve (Swagelok). The mass flow controller (Bronkhorst El-Flow F-201CB) then allows to change the mass flow from 0.04~l/min to 2~l/min. The gas flow is split up and guided through 12 needle valves, which were adjusted individually to reach equal gas flow through each inlet. The gas inlets of the sample cell have an inner diameter of 0.5~mm. This gives expected gas velocities at the inlets of around 7~m/s when using gas streams of around 1~l/min, far below of the speed of sound but above the expected values for the minimal fluidization velocity for sub-millimeter particles in the range of few~cm/s. \cite{Castellanos1999,Durian1997,Ramos1996}. The gas leaves the sample cell through the side panels of the sample cell, which consists of glass frits with 4~$\mu$m pore size.

An additional challenge with gas-fluidization arises with triboelectric charging of the agitated particles in the stream of dry nitrogen used in the experiments. We set a controlled humidity of 75~\% as a countermeasure against triboelectric charging \cite{Ramos1996}. The humidity was set by guiding the gas stream over sponges soaked with oversaturated sodium chloride solution, what precisely set the humidity of the gas stream \cite{Greenspan1977} (see Fig.~\ref{fig:3}). The influece of triboelectric charging on the measured autocorrelation functions was tested by removing the humidity chamber in one drop tower flight.

%--------------------------------------------------------------------------------------------------

\section{Setup and sample cell}

\begin{figure*}
	\centering
	\includegraphics[width=0.75\textwidth]{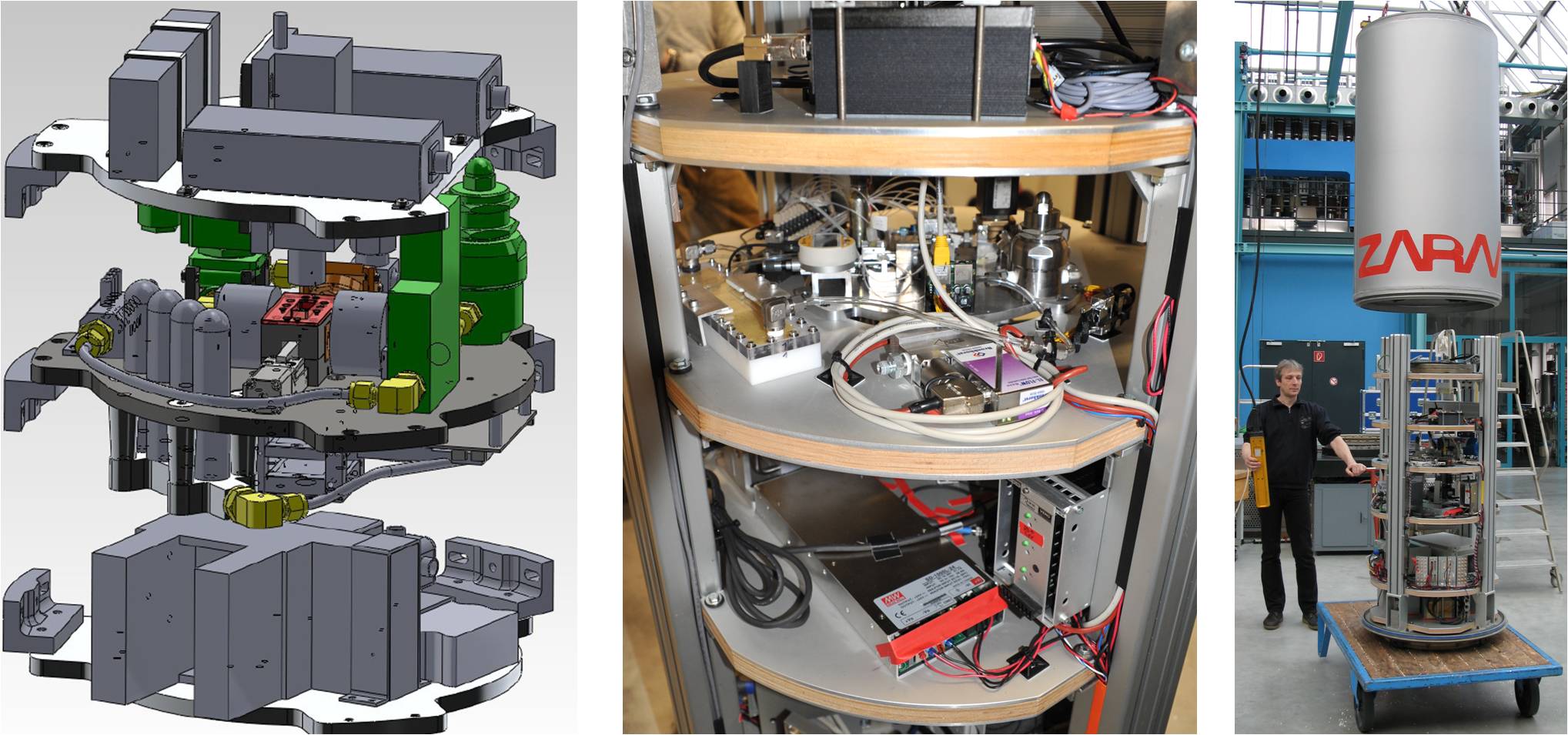}
	\caption{Overview of the three-level design with light source on the top base plate, sample cell with agitation system on the middle base plate and camera and fiber optics on the lower base plate. Left: CAD-drawing of the setup in sounding rocket configuration.  Middle: Close-up of the drop-tower configuration. Right: Integration of the drop tower setup.}
	\label{fig:1} 
\end{figure*}

The setup is designed for experiments on a sounding rocket as well as for drop tower flights.  Thus it is laid out to be autonomous, with only the power supply and an external trigger signal to be provided by the microgravity platform. The setup has to contain the light scattering apparatus as described in section 2, the sample cell providing a variable volume and comprising a gas agitation system as described in section 3, and the data acquisition, the overview camera and the process control.

This is realized by a three-level setup  (see Fig.~\ref{fig:1}). The light scattering apparatus is arranged along the vertical axis (see Fig.~\ref{fig:2}) and the gas agitation system is arranged horizontally on the middle plane (see Fig.~\ref{fig:3}). The light source with power supply is located on the top. It shines light into the sample cell on the middle. The transmitted light is detected by a collimator and fiber under the middle plane, which feeds the detected light into the detection setup on the lowest plane. Additionally, scattered light is monitored by the 500~fps overview camera (Photron Fastcam MC2) on the lowest plane. The setup has a weight of around 25~kg and a height of 40~cm.

The final design of the sample cell consisted of a custom-made aluminum frame with a top and bottom acrylic glass window for measuring in transmission geometry, and a side window for observation with the overview camera (see Fig~\ref{fig:2}). The sample cell has a quadratic cross section of $1\times1$~cm$^{2}$. The length of the sample cell perpendicular to the light transmission path is adjustable by a piston and a precision stepper motor (PI micos VT-21 S) with 50~nm steps from 1~cm to 2~cm. An initial sample cell volume of 1414.5~mm$^{3}$ was filled with 961.9~mm$^{3}$ of sample for the drop tower experiments. The volume fraction of the sample thus was varied between 0.68 (piston position +0~mm) and 0.5 (piston position +5~mm). Piston motion and volume changes are additionally verified by a distance gauge (Sylvac S-Dial NANO). The volume fraction was changed in between drop tower flights and was left unchanged during the measurements. Spherical polystyrene particles with 220~$\mu$m diameter were used for the experiments described in the next section. An overview of the experimental parameters is given in table~\ref{tab:1}

The hardware correlator is connected to a mini-PC (iEi PM-PV-D5251) to store the acquired date. Measurements are started with a time delay after an external trigger signal indicated microgravity. The shut-off valve and the mass flow control are controlled by the National Instrument PXI system of the drop tower capsule in th drop tower experiments. For sounding rocket flights the instruments could be controlled by the mini-PC and a NI compactDAQ. 
\begin{table}%
\centering
	\begin{tabular}{lr}
	\hline\hline
	particle diameter $d$ & 220~$\mu$mm \\
	particle polydispersity & $\approx$5~\% \\
	piston position & 0\ldots 5~mm \\
	cell volume & 1414.5\ldots 1914.5~mm$^{3}$ \\
	particle volume fraction & 0.68\ldots 0.5\\
	particle number density & 12.2\ldots 9~mm$^{-3}$ \\
	gas flow & 0.1\ldots 2~l/min \\
	measurement time & 8~s \\
	pressure$^{\ast}$ & $\approx$1~bar \\
	temperature$^{\ast}$ & $\approx27^{\circ}$C \\
	$l^{\ast}$$^{\dag}$ & $<1d$ \\
	$l_{a}$$^{\dag}$ & $\approx17d$ \\
	rest acceleration$^{\ddag}$ & $10^{-6}$~g$_{0}$ \\
	\hline\hline
	\end{tabular}
	\caption{Overview of the experimental parameters. $^{\ast}$ The drop tower capsule conserved ambient conditions during the drop tower flights. $^{\dag}$ The optical parameters of the samples are preliminary and are based on laboratory transmission measurements presented in ref.~\cite{Reinhold2012}. The extremely low value of the randomization length may be linked to the opaque, white appearance of the polymer particles, thus multiple scattering occurs within single particles. $^{\ddag}$ Confirmed by the drop tower capsule's inertial measurement unit. }
	\label{tab:1}
\end{table}

%--------------------------------------------------------------------------------------------------

\section{Results and Discussion}

Here we show first results on the ability to fluidize the granular medium in microgravity. First we focus on the influence of the strong accelerations during the catapult start, which we analyze using time resolved correlations (TRC) \cite{Cipelletti2003}. Then we compare the influence of the parameters that are adjustable by the setup, i. e. gas humidity, gas flow and piston position, on the measured intensity correlation function.

\subsection{Influence of the catapult start}

\begin{figure}
    \centering
  	\includegraphics[width=0.45\textwidth]{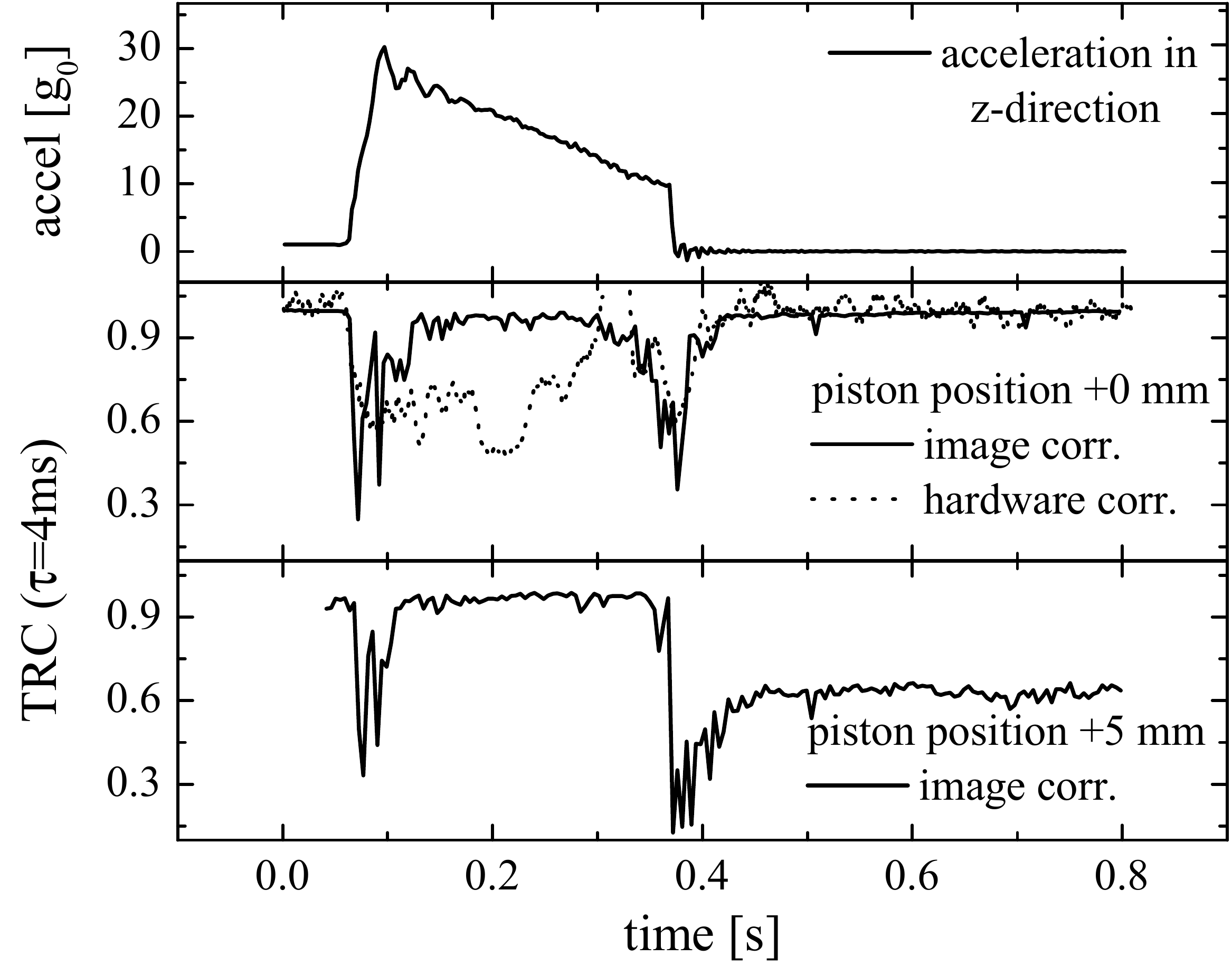}
		\caption{Results of the time resolved correlation (TRC) analysis of the impact caused by the catapult start. The upper diagram shows the acceleration experienced by the setup during the catapult start, in units of the earth gravitational acceleration $g_{0}$. The middle diagram shows the magnitude of correlation with 4~ms delay time, as obtained from image correlation and from the count rate trace of the hardware correlator, as function of time for a jammed sample. No long-lasting vibrations or intensity fluctuations are observable after the catapult is left. The lower diagram shows the result of image correlation as in the middle diagram, but for a diluted and agitated granular medium (2~l/min, compare Fig.~\ref{fig:6}). A steady value of lowered correlation is reached within $\cong$0.2~s after the catapult shot.}
		\label{fig:4}       % Give a unique label
\end{figure}

The overview camera was used to make videos throughout the whole procedure of a drop tower flight, thus the catapult start, the flight and the landing can be monitored. Videos were evaluated by calculating the image correlation between frames with 4~ms delay and averaging the correlation over the whole image. Image correlation were performed with a conventional particle image velocimetry (PIV) code, albeit with zero pixel offset and plotted as a function of time \cite{Pust2000}. The result of this evaluation for catapult starts of a granular medium jammed under compression by the piston and a medium with a retracted piston are shown in Fig.~\ref{fig:4}. On ground, prior to the catapult start, both samples do not show any loss of correlation. When the catapult starts accelerating the capsule, strong loss of correlation sets in, indicating relative motion of the parts of the setup and vibrations. However, when the capsule has left the catapult, the correlation quickly reaches values of unity again, indicating that no vibrations or motions in the setup endure during the microgravity flight. We also used the hardware correlator to track the intensity recorded during this flight. The TRC evaluation of this intensity trace shows the same behavior as observed with the overview camera: after the catapult is left, no intensity fluctuations last and the correlations reaches unity again. This measurements on jammed media shows that no fluctuations are due to the source or the detection, and all measured fluctuations must come from the granular medium.

The lower diagram in Fig.~\ref{fig:4} shows the same image correlation evaluation of a catapult start with a diluted and agitated granular medium. Prior to the catapult start the granular medium stays static, as correlation is not reduced. After the direct influence of the catapult has ceased, the correlation does not reach unity again, but approaches a constant value of 0.6 within $\approx$~0.2~s. The agitation mechanism thus manages to induce a steady state of dynamic particles within a fraction of a second even after the strong compaction of the granular medium during the catapult start. Nevertheless, we set a delay time of 1~s after the capsule has left the catapult to ensure fading of all vibrations and induction of a steady state prior to triggering the hardware correlator.

\subsection{Diffusing wave spectroscopy on fluidized media}

The catapult flights enabled 9~s flights of excellent microgravity conditions (rest accelerations down to $10^{-6}$~g$_{0}$, as confirmed by the drop tower capsule's inertial measurement unit). In all the flights, the measurement time was set to 8~s, with a delay of 1~s after the catapult start, for measurements with the hardware correlator. 

\begin{figure}
    \centering
  	\includegraphics[width=0.4\textwidth]{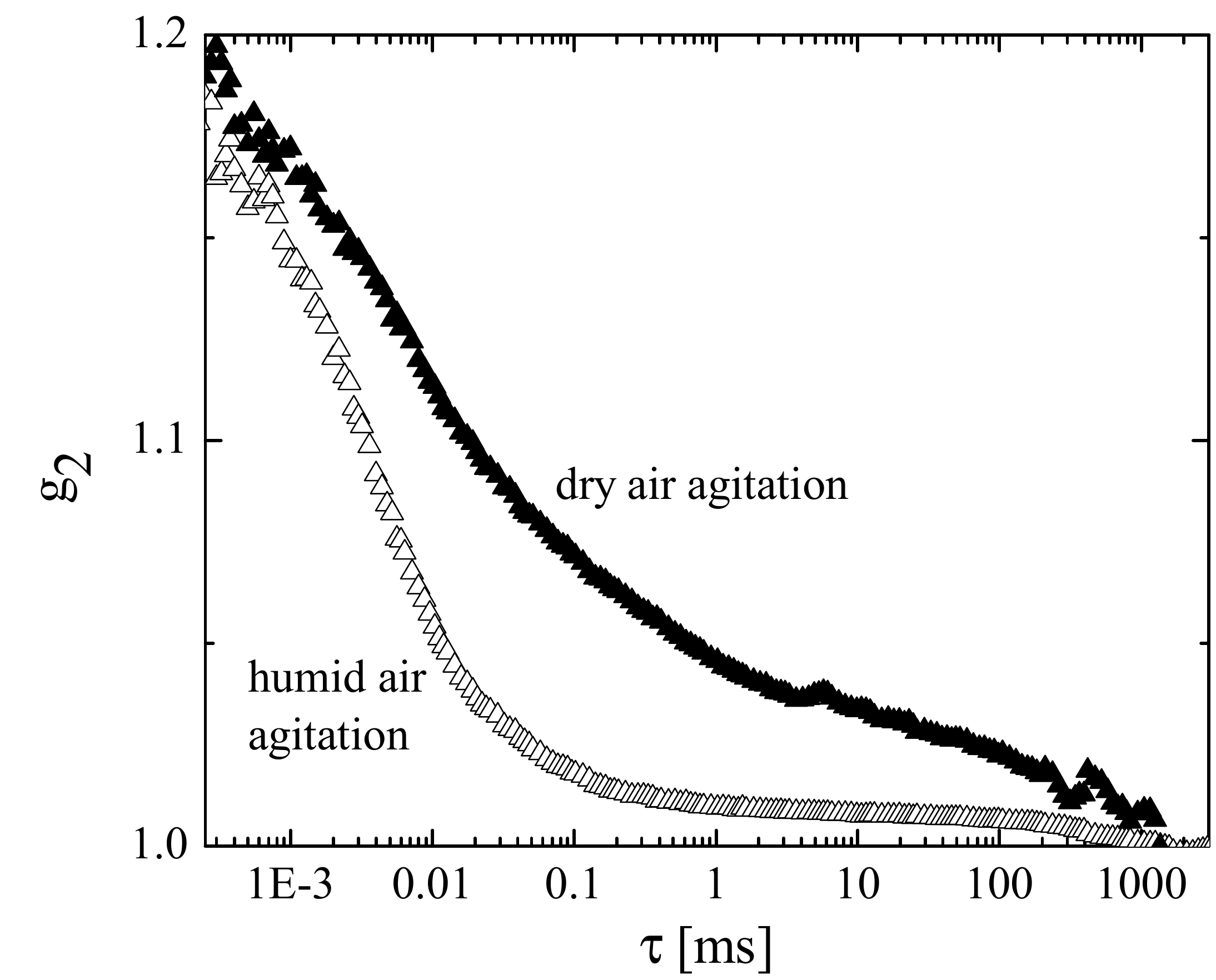}
		\caption{Intensity autocorrelation functions obtained with dry air and humid air agitation in drop tower flights. The agitation with dry air charges the particles and slows down dynamics.}
		\label{fig:5}       % Give a unique label
\end{figure}

\paragraph{Gas humidity.} The humidity of the nitrogen used for fluidizing the granular medium was controllable by introducing or removing the humidity chamber from the gas agitation system (compare Fig.~\ref{fig:3}). The gas humidity was essentially 0~\% without the humidity chamber, and was set to 75~\% by an oversaturated rock salt solution with the humidity chamber. The polystyrene particles considerably charged when agitated with dry air. This became visible by increasing cohesion among the particles and adhesion to the sample cell. The increased stickiness also affected the intensity correlation function (Fig.~\ref{fig:5}). The dynamics was much slower and correlation decay happenes on orders of magnitude longer time scales with dry air than with humid air in drop tower flights with a volume fraction of 0.5 (piston position +5~mm, as before) and 1~l/min gas flow. It seems that in the dry gas case the dynamics was dominated by motion of aggregates rather than the desired collisional motion of individual particles. We consequently used agitation with humid gas in all following measurements.

\begin{figure}
    \centering
  	\includegraphics[width=0.4\textwidth]{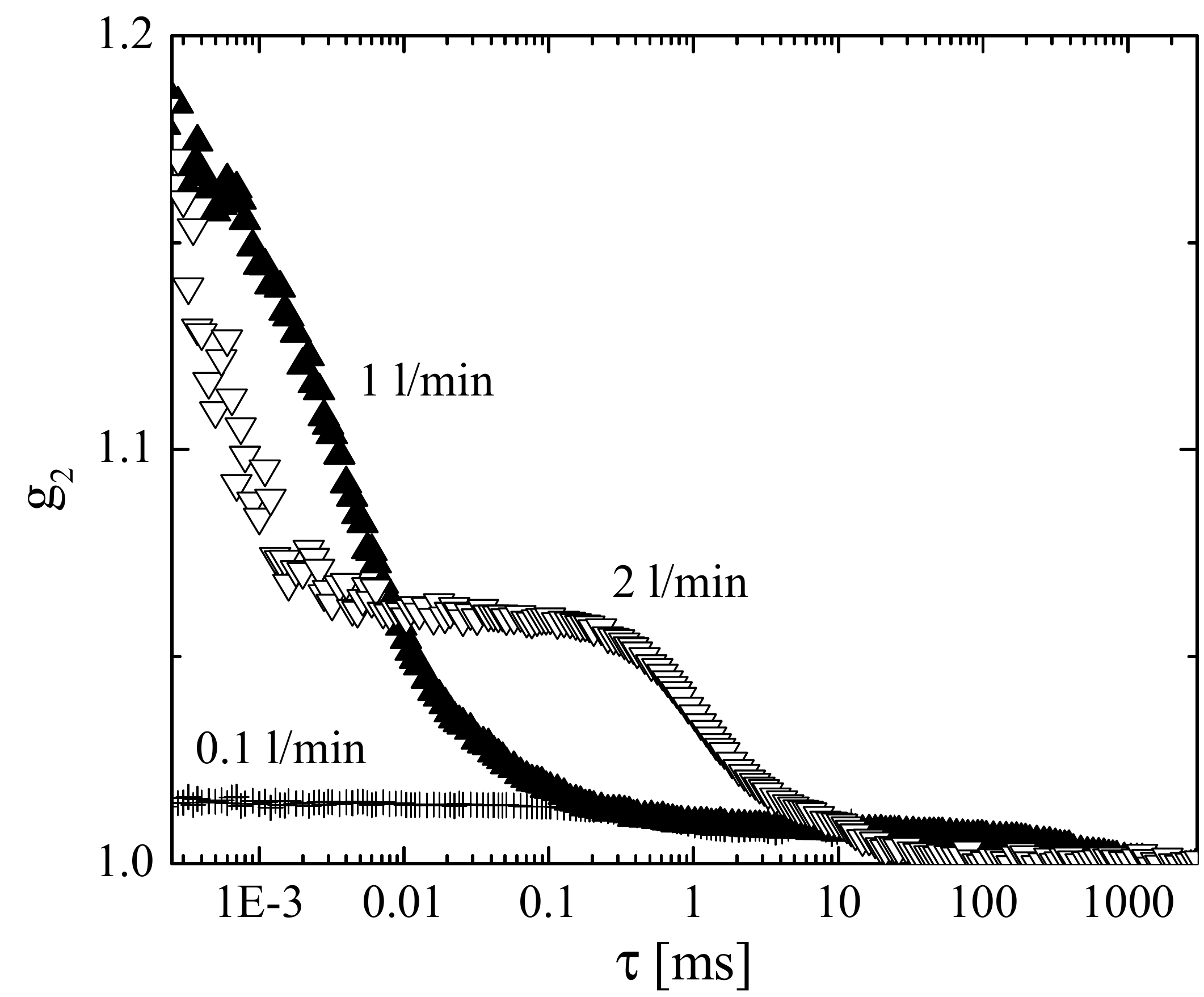}
		\caption{Intensity autocorrelation functions obtained with different gas flows. Increasing the gas flow induces a faster decay of the correlation function and the emergence of a plateau in correlation.}
		\label{fig:6}       % Give a unique label
\end{figure}

\paragraph{Gas flow.} Figure~\ref{fig:6} displays measurements from three drop tower flights with varied gas flow and a particle volume fraction of 0.5 (piston position +5~mm). The gas flow also has a tremendous influence on the shape of the measured correlation functions. Barely any fluctuations and decay in correlation are observable with a gas flow of 0.1~l/min. With a gas flow of 1~l/min a fast decay in correlation within tens of microseconds emerges, followed by a barely visible second decay within hundreds of milliseconds. The intercept of the correlation function with the y-axis at $\approx$1.2 is about half of the reachable intercept without polarizing optics, being 1.5. This indicates that in addition to the unpolarized detection the dynamic contrast is reduced by detection of roughly two independent speckles with the fiber \cite{Durian1999}. The first decay becomes faster when increasing the mass flow to 2~l/min. Additionally a pronounced plateau in correlation emerges, that decays after tens of milliseconds. A plateau in correlation has been observed in previous studies on dense granular media on ground \cite{Goldman2006,Xie2006,Durian1997,Biggs2007}. The plateau had been interpreted as an indication of a localization of the particles, i.e. a restricted motion of the individual particles within cages formed by the respective neighboring particles, which should be observable as a delayed decay in correlation with suitable experimental conditions. The emergence of this plateau with increasing the agitation strength and at a low volume fraction still is remarkable and should motivate further investigation. The measured correlation functions, however, indicate that a fully fluidized granular medium with fast dynamics on microsecond time scales can be induced in microgravity with sufficient gas flow.

\begin{figure}
    \centering
  	\includegraphics[width=0.4\textwidth]{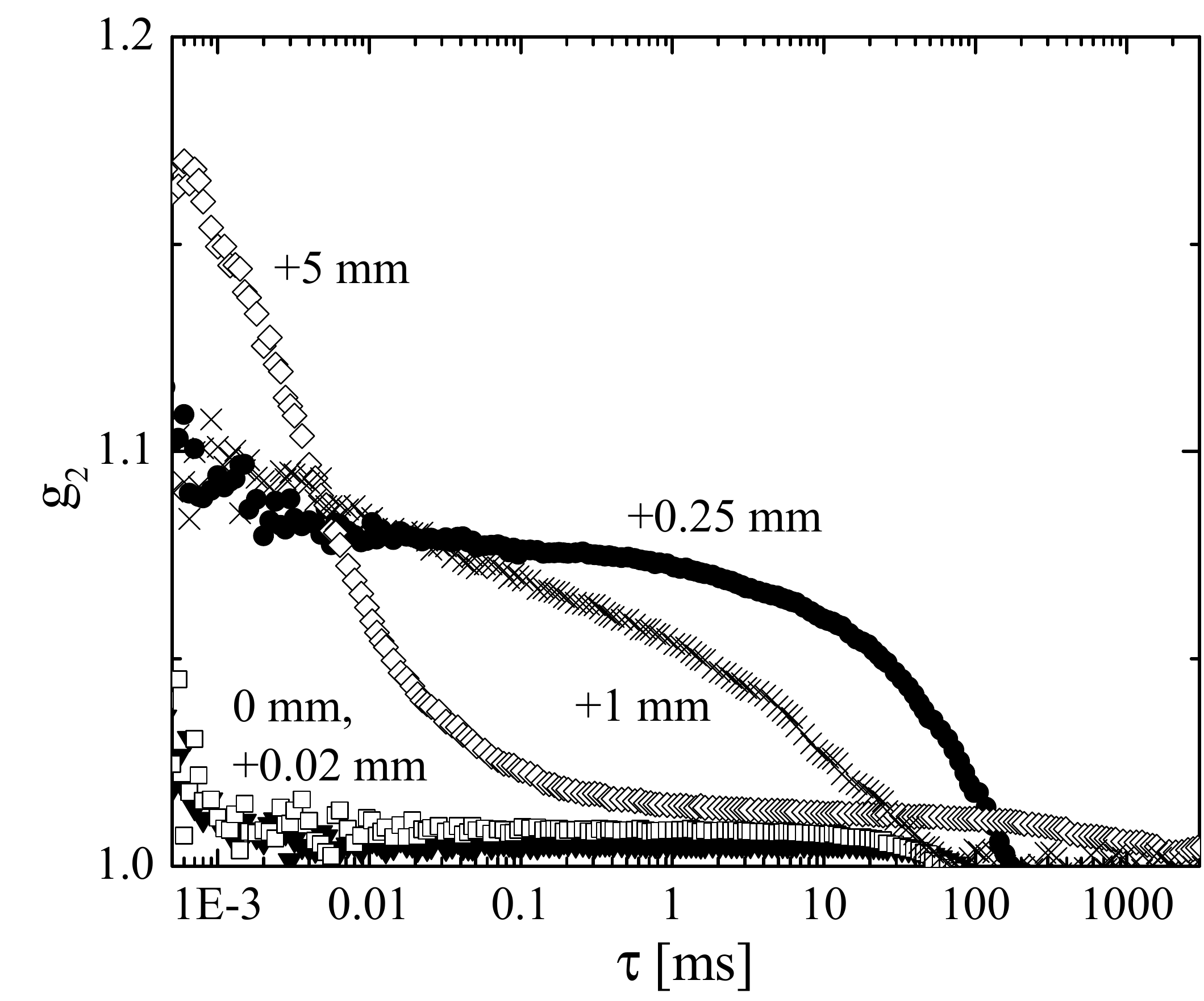}
		\caption{Intensity autocorrelation functions obtained with different piston positions. Moving the piston outwards increases the sample cell volume and lowers the packing fraction in the cell. The sample turns from completely static to fluidized with lower packing fraction.}
		\label{fig:7}       % Give a unique label
\end{figure}

\paragraph{Piston position.} Measurements to test the influence of the piston position and the particle volume fraction with intermediate agitation of 1~l/min were performed as a last series of drop tower flights (Fig.~\ref{fig:7}). The correlation functions do not show any fluctuations and decay in correlation under compression of the piston, what changes only marginally when the piston is retracted by 20~$\mu$m. With larger retraction of the piston more and more fluctuations take place and the intercept of the correlation function increases. The decay in correlation also becomes faster with lower packing fraction. A steep, nearly exponential decay only evolves when the volume of the sample cell is increased by 50~\%. Before that, the correlation functions decay very slowly over several orders of magnitude in time. The reduced intercept and the slow decay in correlation are reminiscent of correlation curves obtained with intermittend dynamics \cite{Durian2001}, and may indicate that either no full fluidization or no steady state is achieved with a continuous gas flow of 1~l/min and high packing fractions. Further work will focus on this issue of granular dynamics.

%--------------------------------------------------------------------------------------------------

\section{Conclusion}

Agitation strength and packing fraction become disentangled in microgravity. Experiments in microgravity thus enable exploring an enlarged parameter space for granular media in comparison to ground based measurements. Dynamics within bulk dense granular media had not been accessible so far in experiments in microgravity. We demonstrate a setup that allows such measurements on dense gas-fluidized granular media with diffusing wave spectroscopy. We can show with this setup that the gas-fluidization enables a steady dynamic state in microgravtiy within fractions of a second. The time scales of intensity fluctuations from such a steady fluidized granular medium are well resolved within the flight time of catapult shots in the drop tower. Triboelectric charging was shown to slow down particle dynamics, but apparently can be effectively minimized by controlled humidity. 

The setup allows controlling gas flow and packing fraction. These parameters induce two noticeable features in the intensity correlation functions, which suggest further work. First, at strong driving and low volume fraction a plateau in correlation can be observed. This behavior is difficult to be interpreted like in ground-based measurements as a localization of the particles. Second, the correlation decays very slowly and far from exponential at higher packing fractions and intermediate driving. This behavior may indicate that particle motion at this higher packing fractions is not steady collisional, but rather dominated by intermittent dynamics. Deeper understanding and refined interpretation of the correlation curves is thus essential.

%--------------------------------------------------------------------------------------------------

\begin{acknowledgements}
The authors thank the team from the ZARM Drop Tower Operation and Service Company (ZARM FAB mbH) for valuable technical support during the finalization of the setup and the measurement campaign. The European Space Agency is acknowledged for providing access to the drop tower by ESA-AO-2009-0943 'Compaction and Sound Transmission in Granular Media'. Financial support by DFG research unit FOR 1394 is gratefully acknowledged. P. B. thanks Andreas Meyer for his continued support of the project.
\end{acknowledgements}

%--------------------------------------------------------------------------------------------------

%--------------------------------------------------------------------------------------------------


\begin{thebibliography}{}
\bibitem{Poschel2000} C. Saluena and T. P\"oschel, \emph{Convection in horizontally shaken granular material}, Eur. Phys. J. E - Soft Matter \textbf{1,} 55 (2000).

\bibitem{Lohse2007} P. Eshuis, K. van der Weele, D. van der Meer, R. Bos, and D. Lohse, \emph{Phase diagram of vertically shaken granular matter}, Phys. Fluids \textbf{19,} 123301 (2007).

\bibitem{Meyer2013} M. Siegl, F. Kargl, F. Scheuerpflug, J. Drescher, C. Neumann, M. Balter, M. Kolbe, M. Sperl, P. Yu, and A. Meyer, \emph{Material Physics Rockets MAPHEUS-3/4: Flights and Developments}, in: Proceedings of the 21st ESA Symposium on European Rocket and Balloon Programmes and Related Research, no. 1, 1 (2013).

\bibitem{Stannarius2013} K. Harth, U. Kornek, T. Trittel, U. Strachauer, S. H\"ome, K. Will, and R. Stannarius, \emph{Granular Gases of Rod-Shaped Grains in Microgravity}, Phys. Rev. Lett. \textbf{110,} 144102 (2013).

\bibitem{Blum2000} J. Blum, G. Wurm, S. Kempf, T. Poppe, H. Klahr, T. Kozasa, M. Rott, T. Henning, J. Dorschner, R. Schr\"apler, H. U. Keller, W. J. Markiewicz, I. Mann, B. A. S. Gustafson, F. Giovane, D. Neuhaus, H. Fechtig, E. Gr\"un, B. Feuerbacher, H. Kochan, L. Ratke, A. El Goresy, G. Morfill, S. J. Weidenschilling, G. Schwehm, K. Metzler, and W.-H. Ip, \emph{Growth and Form of Planetary Seedlings: Results from a Microgravity Aggregation Experiment}, Phys. Rev. Lett. \textbf{85,} 2426 (2000).

\bibitem{Losert2013} N. Murdoch, B. Rozitis, K. Nordstrom, S. F. Green, P. Michel, T.-L. de Lophem, and W. Losert, \emph{Convection in Microgravity}, Phys. Rev. Lett. \textbf{110,} 018307 (2013).

\bibitem{Yu2014} P. Yu, S. Frank-Richter, A. B\"orngen, and M. Sperl, \emph{Monitoring three-dimensional packings in microgravity}, Granul. Matter \textbf{16,} 165 (2014).

\bibitem{Pine1990} D. J. Pine, D. A. Weitz, G. Maret, P. E. Wolf, E. Herbolzheimer, and P. M. Chaikin, \emph{Dynamical Correlations of Multiply Scattered Light}, in: Scattering And Localization Of Classical Waves In Random Media \textbf{8,} 312, P. Sheng (Ed.), World Scientific Publishing Co. Pte. Ltd, Singapore (1990).

\bibitem{Goldman2006} D. Goldman and H. Swinney, \emph{Signatures of Glass Formation in a Fluidized Bed of Hard Spheres}, Phys. Rev. Lett. \textbf{96,} 145702 (2006).

\bibitem{Xie2006} L. Xie, M. J. Biggs, D. Glass, A. S. McLeod, S. U. Egelhaaf, and G. Petekidis, \emph{Granular temperature distribution in a gas fluidized bed of hollow microparticles prior to onset of bubbling}, Europhys. Lett. \textbf{74,} 268 (2006).

\bibitem{Sperl2012} M. Sperl, W. T. Kranz, and A. Zippelius, \emph{Single-particle dynamics in dense granular fluids under driving}, Europhys. Lett. \textbf{98,} 28001 (2012).

\bibitem{Hou2012} Y.-P. Chen, P. Evesque, and M.-Y. Hou, \emph{Breakdown of Energy Equipartition in Vibro-Fluidized Granular Media in Micro-Gravity}, Chinese Phys. Lett. \textbf{29,} 074501 (2012).

\bibitem{Evesque2010} P. Evesque, \emph{Microgravity and Dissipative Granular Gas in a vibrated container: a gas with an asymmetric speed distribution in the vibration direction, but with a null mean speed everywhere}, Poudres \& Grains \textbf{18,} 1 (2010).

\bibitem{Grace2006} J. R. Grace, B. Leckner, J. Zhu, and Y. Cheng, \emph{Fluidized Beds}, in: Multiphase Flow Handbook, pp. 5,1--93, C. T. Crowe (Ed.), CRC Press, (2006).

\bibitem{Ricka1996} W. Leutz and J. Ricka, \emph{On light propagation through glass bead packings}, Opt. Commun. \textbf{126,} 260 (1996).

\bibitem{Castellanos2005} A. Castellanos, \emph{The relationship between attractive interparticle forces and bulk behaviour in dry and uncharged fine powders}, Adv. Phys. \textbf{54,} 263 (2005).

\bibitem{Castellanos1999} A. Castellanos, J. Valverde, A. Perez, A. Ramos, and P. Watson, \emph{Flow Regimes in Fine Cohesive Powders}, Phys. Rev. Lett. \textbf{82,} 1156 (1999).

\bibitem{Durian1997} N. Menon and D. J. Durian, \emph{Particle Motions in a Gas-Fluidized Bed of Sand}, Phys. Rev. Lett. \textbf{79,} 3407 (1997).

\bibitem{Ramos1996} J. Guardiola, V. Rojo, and G. Ramos, \emph{Influence of particle size, fluidization velocity and relative humidity on fluidized bed electrostatics}, J. Electrostat. \textbf{37,} 1 (1996).

\bibitem{Greenspan1977} L. Greenspan, \emph{Humidity fixed points of binary saturated aqueous solutions}, J. Res. Natl. Bur. Stand. Sect. A Phys. Chem. \textbf{81A,} 89 (1977).

\bibitem{Reinhold2012} S. Reinhold, \emph{Lichtstreuung an getriebenen granularen Medien}, Diploma Thesis, Bonn University, 2012.

\bibitem{Cipelletti2003} L. Cipelletti, H. Bissig, V. Trappe, P. Ballesta, and S. Mazoyer, \emph{Time-resolved correlation: a new tool for studying temporally heterogeneous dynamics}, J. Phys. Condens. Matter, \textbf{15,} S257 (2003).

\bibitem{Pust2000} O. Pust, \emph{Direct Cross-Correlation compared with FFT-based Cross-Correlation}, in: Proceedings of the 10th International Symposium on Applications of Laser Techniques to Fluid Mechanics, \textbf{27,} 114 (2000).

\bibitem{Durian1999} P.-A. Lemieux and D. J. Durian, \emph{Investigating non-Gaussian scattering processes by using nth-order intensity correlation functions}, J. Opt. Soc. Am. A \textbf{16,} 1651 (1999).

\bibitem{Biggs2007} M. J. Biggs, D. Glass, L. Xie, V. Zivkovic, A. Buts, and M. A. Curt Kounders, \emph{Granular temperature in a gas fluidized bed}, Granul. Matter \textbf{10,} 63 (2007).

\bibitem{Durian2001} P. A. Lemieux and D. J. Durian, \emph{Quasi-elastic light scattering for intermittent dynamics}, Appl. Opt. \textbf{40,} 3984 (2001).

\end{thebibliography}
\end{document}